\documentclass[12pt]{article}

\usepackage{amsmath}
\usepackage{amsthm}
\usepackage{amssymb}
\usepackage{natbib}
\usepackage{caption}
\usepackage{booktabs}
\usepackage{lipsum}
\usepackage{microtype}

\usepackage{graphicx}

\theoremstyle{definition}

\theoremstyle{plain}

\title{Diversity and Inclusion Index with Networks and Similarity: Analysis and its Application\footnote{This study extends the work presented at the 36th Annual Conference of the Japanese Society for Artificial Intelligence (non-refereed) by \citet{kinjo2022}, adding theoretical research, the introduction of another variable related to diversity and estimation of similarity weights, and a new empirical study, and making significant additions and changes \citep{kinjo2022}.}}

\author{Keita Kinjo\\Kyoritsu Women's University, Faculty of Business}
\date{}

\begin{document}

\maketitle

\begin{abstract}
In recent years, the concepts of ``diversity'' and ``inclusion'' have attracted considerable attention across a range of fields, encompassing both social and biological disciplines. To fully understand these concepts, it is critical to not only examine the number of categories but also the similarities and relationships among them. In this study, I introduce a novel index for diversity and inclusion that considers similarities and network connections. I analyzed the properties of these indices and investigated their mathematical relationships using established measures of diversity and networks. Moreover, I developed a methodology for estimating similarities based on the utility of diversity. I also created a method for visualizing proportions, similarities, and network connections. Finally, I evaluated the correlation with external metrics using real-world data, confirming that both the proposed indices and our index can be effectively utilized. This study contributes to a more nuanced understanding of diversity and inclusion analysis.
\end{abstract}


\newpage

\section{Introduction}
Today, ``diversity'' and ``inclusion'' are attracting attention in various areas of society, with ``equity'' emerging as another important consideration. Scholars have noted the interconnectedness of these concepts with the Sustainable Development Goals (SDGs), highlighting their growing significance \citep{Kioupi2020}. For instance, ethical concerns in artificial intelligence research (AI), a rapidly advancing field, have underscored the importance of human diversity, particularly gender diversity \citep{Leavy2018}. Diversity, originally addressed in various disciplines like biology and knowledge, extends beyond human populations. How is diversity measured? Let me briefly introduce existing diversity definitions, discuss the utilities of ensuring diversity in several domains, outline challenges with the existing definitions, and explain the objectives of this study.

Commonly cited diversity definitions include richness (based on the number of categories in a population), Shannon entropy (derived from the logarithm of category proportions), and the Herfindahl–Hirschman index (calculated using the squared proportions of categories). Additionally, a composite index known as ``true diversity'' (or ``hill number''), characterized by a single parameter, has been proposed \citep{Hill1973, Peet1974, Jost2006}. 

The following is a definition of true diversity:
\begin{equation*}
D_q(p) = 
\begin{cases}
\left( \sum_{i=1}^{n} {p_i}^q \right)^\frac{1}{1-q}, & \text{if } q \neq 1, \\
\left( \prod_{i=1}^{n} {p_i}^{p_i} \right)^{-1}, & \text{if } q = 1.
\end{cases}
\end{equation*}
However when $q = 1$ and $p_i = 0$, ${p_i}^{p_i} := 1$.Here, $n$ is the number of categories, $p_i \in [0, 1]$ is the proportion of category $i$, and $\sum_{i=1}^{n} p_i = 1$. $p = (p_1, \ldots, p_n)^T$ is a column vector. $q \in [0, \infty]$ is the variable defining this index.

This index becomes the following indices depending on the value of $q$:
\begin{equation*}
D_0(p) = |\{i \in \{1, \ldots, n\} : p_i > 0\}|,
\end{equation*}
\begin{equation*}
D_2(p) = \left( \sum_{i=1}^{n} {p_i}^2 \right)^{-1},
\end{equation*}
\begin{equation*}
D_\infty(p) = \left( \max_{i \in \{1, \ldots, n\}}{p_i} \right)^{-1}.
\end{equation*}
Thus, it indirectly includes several indices and concepts, such as the Shannon entropy, richness, Herfindahl–Hirschman index, and Berger–Parker index.

Many studies have shown that ensuring this diversity has utilities in biology, society, and artificial intelligence. In this study, utility means that indices of diversity are related to the performance of the population as a whole.
Specifically, let us consider a scenario with $r$ populations. $Ds = (D_1, \ldots, D_r)$ represents a vector consisting of the diversity $D$ of each population, and $ys = (y_1, \ldots, y_r)$ is a vector reflecting the performance-related values $y$ for each population. Diversity has utility when there is a correlation between these $Ds$ and $ys$. However, it is crucial to note that diversity not only holds utility but also carries an obligatory aspect.

In biology, studies highlight the relationship between species diversity and ecosystem productivity, including biomass, as well as the prevalence of infectious diseases \citep{Gibson2001, Keesing2010}. In the social sciences, diversity among individuals within organizations correlates with productivity \citep{Reagans2001, HorwitzHorwitz2007}. Similarly, in economics—particularly urban economics—diversity of goods is related to preferences for cities \citep{Dixit1977}. In information engineering, it has been noted that the diversity of output results in a recommendation system is important \citep{Kunaver2017}. Moreover, within the realm of AI and machine learning, diversity in data and models is important for improving prediction accuracy \citep{Gong2019}, with methods proposed to ensure fairness by upholding diversity \citep{Mitchell2020}.

Despite the widespread attention to diversity across various fields and its practical utility, existing diversity indices encounter several challenges. One significant issue is that many existing definitions, such as the Hill number, fail to incorporate considerations of similarity or dissimilarity (distance) between categories. To address this gap, several indices have been proposed that integrate similarity and proportion. For instance, \citet{Gibson2001} proposed an index based on the categorical distances (taxonomic distinctness), and \citet{Leinster2021} proposed an index that combines similarity with proportion \citep{Gibson2001, Leinster2012, Leinster2021}. However, these studies encountered limitations by not accounting for interactions between categories and relying on externally provided similarity measures, raising questions about the appropriate choice of similarity metrics.

Many existing indices do not consider interactions between categories, yet these interactions are important for understanding utility. For instance, \citet{Reagans2001} emphasize the significance of both diversity and networks within R\&D-related teams. From an ethical perspective, diversity and inclusiveness are indispensable. While the concept of inclusion is multifaceted, inclusion can be considered as an interaction between categories \citep{Roberson2006,Sherbin2017}. A notable study that addresses such networks and proportions is that of \citet{Morales2021}, who examined diversity across various networks by considering state transitions. However, they do not consider the similarities between categories.

To summarize the above discussion, we break it down into two key questions:
\begin{enumerate}
    \item How can we effectively incorporate both similarity and networks into a comprehensive index?
    \item How should we define and quantify similarity in this context?
\end{enumerate}

In this study, I propose a novel diversity and inclusiveness index that considers both similarities and networks. The properties of these indices are thoroughly analyzed, including investigating the mathematical relationships between the proposed index and existing diversity and network indices. Additionally, I propose a method for estimating similarity based on the utility concept mentioned above. Furthermore, I propose a visualization method to aid in interpreting similarities and networks within the data. Finally, real-world data is utilized to investigate the correlation between the external index $ys$ and the proposed index, validating its usefulness.

Section 2 of this study describes the method and analysis of the proposed index. Section 3 presents the validation using real data, and Section 4 delves into a comprehensive discussion of the findings.

\section{Method}

\subsection{Proposed Index}
As described in Section 1, based on existing studies, including \citet{Jost2006}, \citet{Leinster2021}, and \citet{Morales2021}, I propose a novel diversity and inclusion index that integrates considerations of similarity and network dynamics.

The inputs are \(\{p, Z, E, q\}\). Partially, as defined in Section 1, we assume \(n\) categories. It is also possible to assume that there is no population and consider one sample as one category. \(i, j\) are the identification numbers assigned to the categories and have values from 1 to \(n\). \(p = (p_1, \ldots, p_n)^T\) is a vector of \(n\) proportions \(p_i \in [0, 1]\) and \(\sum_{i=1}^{n} p_i = 1\). \(Z = (Z_{i,j})_{1 \le i,j \le n} \in [0,1]\) is an \(n \times n\) similarity matrix. The specific similarities are as follows: Each category \(i\) or \(j\) has an \(a\)-dimensional attribute vector \(x_i, x_j \in [0,1]\). \(L\) is an \(n \times n\) matrix with all elements of 1. \(\bar{Z} = L - Z\) is the dissimilarity matrix (distance matrix). \(E = (E_{i,j})_{1 \le i,j \le n} \in [0,1]\) is the \(n \times n\) adjacency matrix representing the network. The graph is directed. The diagonal elements can also be set to 1, and weights can also be considered. \(q \in [0, \infty]\) is a variable that specifies the type of diversity. The diversity measure for the entire population was as follows:

\textbf{Definition:} Diversity and inclusion index with similarity and network (DSN) is defined as follows:

\begin{equation*}
D_q^{\bar{Z}}(p, E) = \begin{cases}
\left( \sum_{i=1}^{n} p_i \left( \left( L - \bar{Z} \circ E \right) p \right)_i^{q-1} \right)^\frac{1}{1-q}, & \text{if } q \neq 1, \infty, \\
\prod_{i=1}^{n} \left( \left( L - \bar{Z} \circ E \right) p \right)_i^{-p_i}, & \text{if } q = 1, \\
\left( \max_{i \in \{1, \ldots, n\}} \left( (L - \bar{Z} \circ E) p \right)_i \right)^{-1}, & \text{if } q = \infty,
\end{cases}
\end{equation*}
where \(\left( (L - \bar{Z} \circ E) p \right)_i = \sum_{j=1}^{n} (1 - \bar{Z}_{i,j} E_{i,j}) p_j\). \(\circ\) is the Hadamard product. If \(q = 1\), \(\left( (L - \bar{Z} \circ E) p \right)_i = 0\) and \(p_i = 0\), then \(\left( (L - \bar{Z} \circ E) p \right)_i^{-p_i} = 1\).

Examples of similarity matrices \(Z\) are \(Z_{ij} = e^{-d(x_i, x_j)}\) or \(Z_{ij} = \frac{1}{1 + d(x_i, x_j)}\), where \(d\) is the distance function using the attribute vectors \(x_i\) and \(x_j\). If the distance function is such that each value falls within 0–1, it is possible to use its values directly in the dissimilarity matrix \(\bar{Z}\). There are also cases where \(i\) and \(j\) have only one set of attributes. In such cases, the similarities between these sets (the Jaccard coefficient, Dice coefficient, Simpson coefficient, etc.) can be used.

Network \(E\) specifically represents the direction and degree of communication and the degree of relationship in a human organization. In addition to this, \(E(n,\rho) = E + \rho E^2 + \rho^2 E^3 + \ldots + \rho^{n-1} E^n = E + E(\rho E) + (E(\rho E))^2 + \ldots + (E(\rho E))^{n-1} = E \frac{1 - (\rho E)^n}{1 - \rho E}\), which takes the effect of the \(n\)-squared network into consideration. Where \(\rho \in [0,1]\) is the discount rate.

By expressing this as a single index, the interaction between the network and similarity can be considered. Utility can be easily assessed by estimating the correlation or regression between the vector of the diversity index \(Ds_q = (D_q^{\bar{Z}}(p, E)_1, \ldots, D_q^{\bar{Z}}(p, E)_r)\) and the vector of index-related performance \(ys = (y_1, \ldots, y_r)\) in population \(r\). It encompasses several indices. This point is discussed in Section 2.2.

\subsection{Analysis of Proposed Index}

In this section, we discuss the properties of the proposed index. Specifically, we investigate how the differences in \(q\), \(Z\), and \(E\) affect the proposed index. In addition, the relationship between existing and proposed indices is discussed. First, the following proposition holds:

\textbf{Proposition 1.}
\textit{\(D_q^{\bar{Z}}(p, E)\) is monotonically decreasing with respect to \(q\).}

\begin{proof}
Assuming \(\sum_{i=1}^{n} p_i = 1\), \(z_i \in \mathbb{R}\), \(n \in \mathbb{Z}\), and \(f: X \rightarrow \mathbb{R}\) is a convex function, from Jensen's inequality, we have \(f\left(\sum_{i=1}^{n} p_i z_i\right) \le \sum_{i=1}^{n} p_i f(z_i)\). If \(y_i \geq 0\), \(b > 0\), \(c > 0\), \(b < c\), \(f(y_i) = y_i^{\frac{c}{b}}\), \(z_i = x_i^b\), then \(f'(y_i) > 0\) and \(f''(y_i) > 0\). Substituting into the previous inequality, we get
\[
\left( \sum_{i=1}^{n} p_i x_i^b \right)^{\frac{c}{b}} \le \sum_{i=1}^{n} p_i x_i^c \Longleftrightarrow \left( \sum_{i=1}^{n} p_i x_i^b \right)^{\frac{1}{b}} \le \left( \sum_{i=1}^{n} p_i x_i^c \right)^{\frac{1}{c}}.
\]
Since \(\left( \sum_{i=1}^{n} p_i x_i^t \right)^{\frac{1}{t}}\) monotonically increases with respect to \(t\), assuming \(t = 1 - q\), \(x_i = 1 / \left( \left( L - \bar{Z} \circ E \right) p \right)_i\), we have
\[
\left( \sum_{i=1}^{n} p_i \left( \left( L - \bar{Z} \circ E \right) p \right)_i^{-1(1-q)} \right)^{\frac{1}{1 - q}} = \left( \sum_{i=1}^{n} p_i \left( \left( L - \bar{Z} \circ E \right) p \right)_i^{q-1} \right)^{\frac{1}{1 - q}}
\]
monotonically increases with respect to \(1 - q\). That is, it monotonically decreases with respect to \(q\). 
\end{proof}

Next, when \(Z_{i,j} \geq Z'_{i,j}\) for all \(i, j\), we denote in this paper \(Z \geq Z'\). When \(E_{i,j} \geq E'_{i,j}\) for all \(i, j\), we denote \(E \geq E'\). The following proposition holds:

\textbf{Proposition 2.} 
\textit{If \(E\) and \(p\) are fixed, and \(Z \geq Z'\), then \(D_q^{\bar{Z}}(p, E) \le D_q^{\bar{Z}'}(p, E)\). If \(Z\) and \(p\) are fixed, and \(E \le E'\), then \(D_q^{\bar{Z}}(p, E) \le D_q^{\bar{Z}}(p, E')\).}

\begin{proof}
When \(Z \geq Z'\), then \(\bar{Z} \le \bar{Z'}\). \(E\) is fixed, then \(L - \bar{Z} \circ E > L - \bar{Z'} \circ E\). Assuming that \(p\) is fixed, if \(0 \le q < 1\), then \(q - 1 < 0\), so 
\[
p_i \left( \left( L - \bar{Z} \circ E \right) p \right)_i^{q-1} \le p_i \left( \left( L - \bar{Z'} \circ E \right) p \right)_i^{q-1}.
\]
In addition, \(\frac{1}{1 - q} > 0\), we get
\[
\left( \sum_{i=1}^{n} p_i \left( \left( L - \bar{Z} \circ E \right) p \right)_i^{q-1} \right)^{\frac{1}{1 - q}} \le \left( \sum_{i=1}^{n} p_i \left( \left( L - \bar{Z'} \circ E \right) p \right)_i^{q-1} \right)^{\frac{1}{1 - q}}.
\]
If \(q = 1\), then \(-p_i < 0\), so 
\[
\left( \left( L - \bar{Z} \circ E \right) p \right)_i^{-p_i} \le \left( \left( L - \bar{Z'} \circ E \right) p \right)_i^{-p_i}.
\]
We get
\[
\prod_{i=1}^{n} \left( \left( L - \bar{Z} \circ E \right) p \right)_i^{-p_i} \le \prod_{i=1}^{n} \left( \left( L - \bar{Z'} \circ E \right) p \right)_i^{-p_i}.
\]
If \(1 < q < \infty\), then \(q - 1 > 0\),
\[
p_i \left( \left( L - \bar{Z} \circ E \right) p \right)_i^{q-1} \ge p_i \left( \left( L - \bar{Z'} \circ E \right) p \right)_i^{q-1}.
\]
In addition, \(\frac{1}{1 - q} < 0\), we get
\[
\left( \sum_{i=1}^{n} p_i \left( \left( L - \bar{Z} \circ E \right) p \right)_i^{q-1} \right)^{\frac{1}{1 - q}} \le \left( \sum_{i=1}^{n} p_i \left( \left( L - \bar{Z'} \circ E \right) p \right)_i^{q-1} \right)^{\frac{1}{1 - q}}.
\]
If \(q = \infty\), \(\max_{i \in \{1, \ldots, n\}} \left( (L - \bar{Z} \circ E) p \right)_i \ge \max_{i \in \{1, \ldots, n\}} \left( (L - \bar{Z'}) \circ E) p \right)_i\), then
\[
\frac{1}{\max_{i \in \{1, \ldots, n\}} \left( (L - \bar{Z} \circ E) p \right)_i} \le \frac{1}{\max_{i \in \{1, \ldots, n\}} \left( (L - \bar{Z'} \circ E) p \right)_i}.
\]
The above results show that \(D_q^{\bar{Z}}(p, E) \le D_q^{\bar{Z}'}(p, E)\) for all \(q\). 
If \(E \le E'\) and \(Z\) is fixed, \(L - \bar{Z} \circ E \ge L - \bar{Z} \circ E'\). As above, \(L - \bar{Z} \circ E > L - \bar{Z'} \circ E\), \(D_q^{\bar{Z}}(p, E) \le D_q^{\bar{Z}}(p, E')\) for all \(q\). 
\end{proof}

Characteristically, when \(\bar{Z}\) and \(E\) are both higher, the value of \(L - (\bar{Z} \circ E)\) is smaller, thus the diversity measure is basically higher (note the sign of \(q - 1\) or \(\frac{1}{1 - q}\)). In other words, the higher the dissimilarity between categories and the more networks there are, the higher the rating. Next, cases with high similarity and many networks, or dissimilarity and few networks, were evaluated. Finally, cases with high similarity and few networks were rated lower.

I discuss the relationship between the proposed index and various existing diversity- and network-related indices \citep{Jost2006}. This confirms that the proposed index encompasses several indices and the conditions under which they are valid. First, the following can be said about the index of true diversity (or Hill number) \(D_q(p)\) described in Section 1 and the proposed index.

\textbf{Proposition 3.1.} 
\textit{If \(Z\) is the identity matrix \(I\) and \(E = L\), then \(D_q^{\bar{Z}}(p, E) = D_q(p)\).}

\begin{proof}When the above condition and \(q \neq 1, \infty\),
\[
D_q^{\bar{Z}}(p, E) = \left( \sum_{i=1}^{n} p_i \left( (Ip)_i \right)^{q-1} \right)^{\frac{1}{1-q}} = \left( \sum_{i=1}^{n} p_i p_i^{q-1} \right)^{\frac{1}{1-q}} = \left( \sum_{i=1}^{n} p_i^q \right)^{\frac{1}{1-q}}.
\]
They also hold for \(D_q^{\bar{Z}}(p, E)\) when \(q = 1, \infty\). 
\end{proof}
             
I also discuss its relationship with the \(D_q^Z(p)\) index by Leinster (2021), which considers similarity and proportion \citep{Leinster2021}. This index is defined as follows: 
\[
D_q^Z(p) = \begin{cases}
\left( \sum_{i=1}^{n} p_i \left( (Zp)_i \right)^{q-1} \right)^{\frac{1}{1-q}}, & \text{if } q \neq 1, \infty, \\
\prod_{i=1}^{n} \left( (Zp)_i \right)^{-p_i}, & \text{if } q = 1, \\
\left(\max_{i \in \{1, \ldots, n\}} (Zp)_i \right)^{-1}, & \text{if } q = \infty.
\end{cases}
\]

The following holds between this index and the proposed index.

\textbf{Proposition 3.2.} 
\textit{If all elements of \(E\) are 1, then \(D_q^{\bar{Z}}(p, E) = D_q^Z(p)\).}

\begin{proof} When the above condition and \(q \neq 1, \infty\), then
\[
L - \bar{Z} \circ E = L - (L - Z) \circ L = Z,
\]
we get \(D_q^{\bar{Z}}(p, E) = D_q^Z(p)\). They also hold for \(D_q^{\bar{Z}}(p, E)\) when \(q = 1, \infty\). 
\end{proof}

Next, I discuss the relationship between the proposed index and the indices used in graph-based analyses, such as social network analysis. There is an index called network density that is often used in network analysis \citep{WassermanFaust1994, KnokeYang2019}. Let \(Nd = \frac{\sum_{i=1}^{n} \sum_{j=1}^{n} E_{i,j}}{n^2}\) be the network density of a directed graph, including self-loops. The following holds true for this index.

\textbf{Proposition 3.3.} 
\textit{If \(p_i = \frac{1}{n}\), \(Z = 0\), \(q = 2\), then \(D_q^{\bar{Z}}(p, E) = \left(1 - Nd\right)^{-1}\).}

\begin{proof}
\[
D_q^{\bar{Z}}(p, E) = \left( \sum_{i=1}^{n} p_i \left( (L - \bar{Z} \circ E)p \right)_i^{q-1} \right)^{\frac{1}{1-q}} = \left( \sum_{i=1}^{n} \frac{1}{n} \left( (L - E)p \right)_i \right)^{-1}
\]
\[
= \left( \sum_{i=1}^{n} \frac{1}{n^2} \left( \sum_{j=1}^{n} (1 - E_{i,j}) \right) \right)^{-1} = \left( \sum_{i=1}^{n} \left( \frac{1}{n} - \frac{\sum_{j=1}^{n} E_{i,j}}{n^2} \right) \right)^{-1}
\]
\[
= \left( 1 - \frac{\sum_{i=1}^{n} \sum_{j=1}^{n} E_{i,j}}{n^2} \right)^{-1} = \left( 1 - Nd \right)^{-1}.
\]
\end{proof}

The results show that, compared to the usual network density, the proposed index considers additional information, such as the proportion of each category and the degree of similarity. Several studies have used variance to define diversity \citep{PatilTaillie1982}. For example, some studies have defined beta diversity based on the variance in the number of units within each category \citep{AndersonEtAl2006, LegendreDeCaceres2013, ChaoChiu2016, Ricotta2017}. Here, I demonstrate that the proposed or Leinster's index and the variance of attributes are related \citep{Leinster2021}. First, assume that the attributes are one-dimensional values \(x_i, x_j \in [0, 1]\) and define their sample variance as \(S = \frac{1}{n} \sum_{i=1}^{n} (x_i - \bar{x})^2\). The following proposition holds.

\textbf{Proposition 3.4.} 
\textit{Assuming each category is an individual and the dissimilarity \(\bar{Z}_{i,j} = (x_i - x_j)^2\), when \(p_i = \frac{1}{n}\), \(q = 2\), \(E = L\), then \(D_q^{\bar{Z}}(p, E) = \left(1 - S\right)^{-1}\).}

\begin{proof}
\[
D_q^{\bar{Z}}(p, E) = \left( \sum_{i=1}^{n} p_i \left( (L - \bar{Z} \circ E)p \right)_i^{q-1} \right)^{\frac{1}{1-q}} = \left( \sum_{i=1}^{n} \frac{1}{n} \left( (L - \bar{Z})p \right)_i \right)^{-1}
\]
\[
= \left( \sum_{i=1}^{n} \frac{1}{n^2} \left( (1 - (x_i - x_1)^2) + (1 - (x_i - x_2)^2) + \ldots + (1 - (x_i - x_n)^2) \right) \right)^{-1}
\]
\[
= \left( \sum_{i=1}^{n} \frac{1}{n^2} \left( n - \sum_{j=1}^{n} (x_i - x_j)^2 \right) \right)^{-1} = \left( \sum_{i=1}^{n} \left( \frac{1}{n} - \frac{1}{n^2} \sum_{j=1}^{n} (x_i - x_j)^2 \right) \right)^{-1}
\]
\[
= \left( 1 - \frac{1}{n^2} \sum_{i=1}^{n} \sum_{j=1}^{n} (x_i - x_j)^2 \right)^{-1}.
\]
Using Lemma 1 (Appendix),
\[
\left( 1 - \frac{1}{n} \sum_{i=1}^{n} (x_i - \bar{x})^2 \right)^{-1} = \left( 1 - S \right)^{-1}.
\]
\end{proof}

In this setting, the higher the variance of the attributes, the higher the diversity. It is clear that by setting the dissimilarity to the square of the difference and \(E\) to \(L\), a general statistic is encompassed in the proposed index.

\subsection{Methods for Estimating Similarity}

In the previous section, \(Z\) and \(E\) simply used predefined values. Similarities have also been exogenously reported in existing studies \citep{Leinster2021}. However, a question arises as to which similarities or dissimilarities should be used. In addition, the question of which attributes should be emphasized when using the Euclidean distances between attributes remains. This is important because it also relates to the question of whether the diversity of demographic attributes, such as gender and age, or the diversity of tasks, such as skills, within an organization should be emphasized in fields such as management studies \citep{HorwitzHorwitz2007}. Therefore, we propose a method for estimating these similarities based on the utility described in Section 1.

The input is \(\{ys, Ds_q\}\), where \(Ds_q = (D_q^{\bar{Z}}(p, E)_1, \ldots, D_q^{\bar{Z}}(p, E)_r)\). Note that \(n\) and \(\bar{Z}\) were fixed for all populations.

Specifically, I defined a weighted Euclidean distance using weights \(w^* = \{w_1^*, \ldots, w_a^*\}\) for \(x\) such that the correlation is maximized as follows: Finally, the weighted distance was used to calculate diversity.

\[
w^* = \operatorname*{argmax}_{w_1, \ldots, w_a} \textit{cor}(ys, Ds_q)^2,
\]

\begin{center}
subject to \(w_1, \ldots, w_a \in [0, 1]\), \(\sum_{m=1}^{a} w_m = 1\),
\end{center}

where \(\bar{Z} = L - Z\), \(Z_{ij} = e^{-d(x_i, x_j)},\) \(d(x_i, x_j) = \sqrt{(x_i - x_j)^T W (x_i - x_j)},\)
\(W\) is a diagonal matrix where \(w_1, \ldots, w_a\) are the diagonal components, and \(\textit{cor}\) is the correlation function between \(ys\) and \(Ds_q\). Pearson's correlation, Spearman's correlation, and nonlinear correlation (Maximal Information Coefficient; MIC) are available for correlation \citep{ReshefEtAl2011}.

The problem is a complex, constrained nonlinear optimization using weighted Euclidean distances. Therefore, it is necessary to use optimization methods such as sequential least squares programming (SLSQP) \citep{Biggs1975, GillEtAl2019}. It is noteworthy that the optimal solution may depend on the initial values.

\subsection{Visualization}

So far, the research has primarily focused on developing the proposed diversity index. However, the data used to calculate this index consists of several elements, including category proportions, similarities, and networks. The complexity of this can make it challenging to intuitively grasp the characteristics of a group. Therefore, a visualization method is proposed to facilitate the analysis and interpretation of the original data.

The input is \(\{p, E, Z\}\), which is part of the variable. The following procedure was used:
\begin{enumerate}
    \item Centralize the dissimilarity matrix \(\bar{Z}\):
    \[
    \bar{Z}_{ce} = -\frac{1}{2}C\bar{Z}^2C, \quad C = I - \frac{1}{2}L,
    \]
    where \(I\) is the identity matrix.
    
    \item Eigenvalue decomposition of \(\bar{Z}_{ce}\). Then we obtain the matrix \(\mathrm{\Lambda}\) with the eigenvalues \(\lambda\) as diagonal components and matrix \(V\) of the eigenvector \(\vec{v}\):
    \[
    \bar{Z}_{ce} = V\mathrm{\Lambda}V^{-1},
    \]
    where
    \[
    \mathrm{\Lambda} = \begin{pmatrix}
    \lambda_1 & 0 & \ldots \\
    0 & \lambda_2 & \ldots \\
    \ldots & \ldots & \ldots \\
    \end{pmatrix}, \quad V = (\vec{v_1}, \vec{v_2}, \ldots).
    \]

    \item Eigenvalues other than \(\lambda_1, \lambda_2\) of \(\mathrm{\Lambda}\) are set to 0. The following calculations were performed:
    \[
    X^* = V\mathrm{\Lambda}^{\frac{1}{2}}.
    \]
    
    \item Place each category using \(X^*\).
    
    \item The diameter of each circle was calculated based on the values of \(p\) in each category.
    
    \item \(E\) was used to create a network (link) between each category, and the thickness of the network (link) was the width of the network (link).
\end{enumerate}

First, based on \(\bar{Z}\), each category was placed in two dimensions using the multidimensional scaling (MDS) concept \citep{SaeedEtAl2018}. The dissimilarity between categories was preserved in two dimensions. The proportions of the categories were then expressed in terms of the diameter of the circle. Finally, the relationship between categories was represented by a directed graph, with the thickness of the arrows representing the strength of the relationship.

The significance of this visualization is fourfold.(1)The similarities and biases between categories can be understood.
(2) The size and bias of the category distribution can be understood.(3) When dissimilarity and \(E\) are large, there is a distance between categories in two dimensions, and the arrows between them are thicker and therefore appear larger. Therefore, networks that significantly influenced diversity could be identified.(4) Conversely, it is easy to consider the type of network required between the categories.

\section{Empirical Analysis}

\subsection{Data}

To verify the utility of the proposed index, the diversity that people prefer in organizations was investigated. This preference was then compared with several diversity indices to determine whether there was a correlation with the proposed index. If the correlation is positive, the proposed index can be used to design an ethically preferred organization.

\begin{figure}[h]
    \centering
    \includegraphics[width=0.8\textwidth]{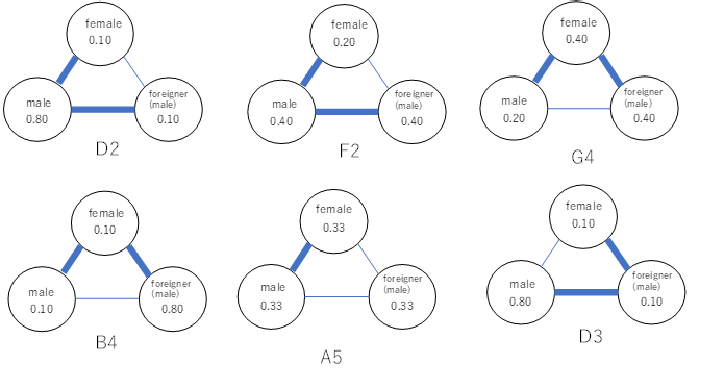}
    \caption{Example of a questionnaire presented}
    \label{fig:questionnaire}
\end{figure}

The details of this study are as follows: First, participants were presented with several hypothetical organizations in Japan. These organizations included Japanese men, Japanese women, and foreigners (men), with each category's proportion and the degree of communication between categories represented by the thickness of network links in an undirected graph (representing bi-directional connections) (Figure \ref{fig:questionnaire}). Next, the study investigated people's preferences for such organizations using a specific question: ``Imagine a Japanese company where an organization has the following proportions and connections: Please rate this organization on an 11-point scale, with 10 being `good,' 5 being `undecided,' and 0 being `not good'. The number represents the proportion of individuals, and the network represents the degree of communication (more or less).''

There are seven different settings for the proportions of Japanese males, Japanese females, and foreigners (males), as follows:(0.33, 0.33, 0.33), (0.80, 0.10, 0.10), (0.10, 0.80, 0.10), (0.10, 0.10, 0.80), (0.20, 0.40, 0.40), (0.40, 0.20, 0.40), (0.40, 0.40, 0.20).

There are two types of network (communication) between each category: ``strong'' and ``weak.'' There were three networks; therefore, in total, there were \(2^3 = 8\) types. Based on the above, the proportions and networks of all populations were \(7 \times 8 = 56\) types (each named A1 to G8). As it was difficult to ask each individual all 56 types of questions, a questionnaire with ten patterns was prepared, consisting of six types \(\times\) six patterns and five types \(\times\)four patterns randomly selected from the 56 types. These were randomly presented to 85 Japanese university students (females) aged 19 and 20 who were asked to answer the questions. The survey was conducted online in November 2021.

Table 1 presents the descriptive statistics for the preferences. The mean of the 56 types has an approximate mean of 5.256 and a standard deviation of 1.362.

\captionsetup[figure]{labelformat=empty, position=above}
\begin{figure}[!htbp]
    \centering
    \caption{Table 1: Descriptive statistics}
    \includegraphics[width=0.4\textwidth]{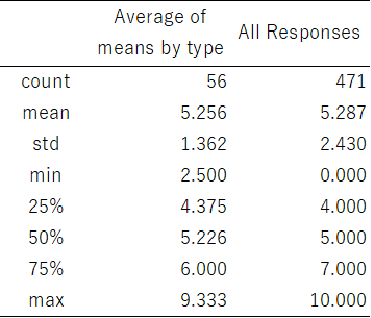}
    \label{Table1:descriptive_stats}
\end{figure}

\subsection{Result}

To calculate the similarity, I optimized the weight vector \(w^*\) to achieve high correlation (Pearson), ordinal correlation (Spearman), and nonlinear correlation (using MIC). The specific settings used are as follows: (1) I utilized a two-dimensional attribute vector \(x\), where gender and origin were expressed as 0-1. Specifically, Japanese women, Japanese men, and foreigners (men) were represented as \(\left(1,0\right)\), \(\left(1,1\right)\), and \((0,0)\), respectively, based on their gender and origin.
    (2) \(w^*\) was calculated as described in Section 2.3 using the SLSQP optimization method, with the weighted Euclidean distance as the distance function \(d\left(x_i, x_j\right)\). \(E_{i,j}\) was set to 1 for more communication and 0.5 for less communication. 

Table 2 presents the results for \(w^*\) obtained by varying \(q\) values from 0 to 0.5, 1, 2, and 10. The results below show the importance of attributes with different \(q\) values, although there was no significant difference observed between the types of correlations.

\begin{figure}[!htbp]
    \centering
    \captionsetup{position=above}
    \caption{Table 2: Optimal values for correlations, ordinal correlations, and nonlinear correlations}
    \includegraphics[width=0.7\textwidth]{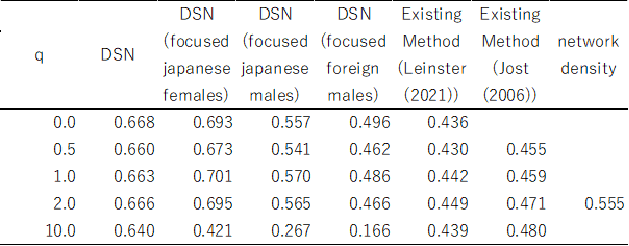} 
    \label{fig:table2}
\end{figure}

Correlations, rank correlations, and nonlinear correlations were calculated and compared among diversity indices, where all elements of \(E\) are 1 indices, using the adjacency matrix of a directed graph centered on Japanese women, Japanese men, and foreign men (with 0 in all but the rows corresponding to the focused category), along with the mean values of preferences (Tables 3-1, 3-2, 3-3).

\begin{figure}[!htbp]
    \centering
    \captionsetup{position=above}
    \caption{Table 3-1: Correlation}
    \includegraphics[width=0.7\textwidth]{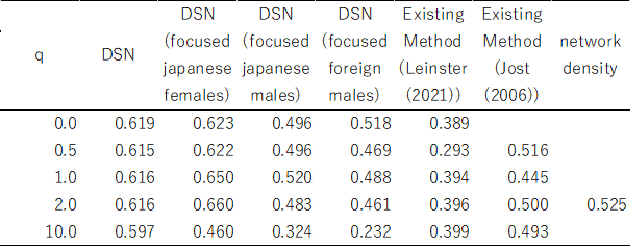} 
    \label{fig:table3-1}
\end{figure}

\begin{figure}[!htbp]
    \centering
    \captionsetup{position=above}
    \caption{Table 3-2: Ordinal correlation}
    \includegraphics[width=0.7\textwidth]{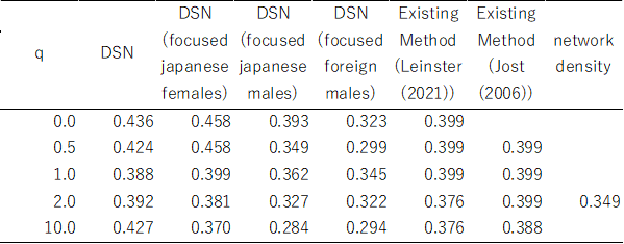} 
    \label{fig:table3-2}
\end{figure}

\begin{figure}[!htbp]
    \centering
    \captionsetup{position=above}
    \caption{Table 3-3: Nonlinear correlation}
    \includegraphics[width=0.7\textwidth]{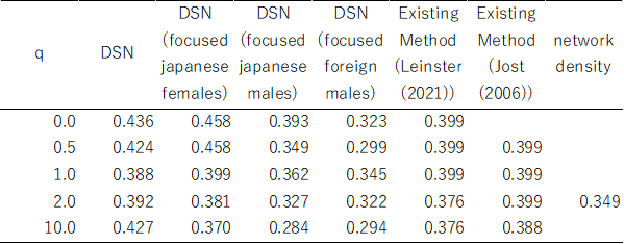} 
    \label{fig:table3-3}
\end{figure}

The three correlations between DSNs and preferences are higher compared to the correlations between existing indices, as well as network density, and preferences. This suggests that DSNs more accurately represent preferences. This finding suggests the need to examine diversity and inclusiveness by considering similarities and networks.

Finally, visualization experiments were conducted using similarity with optimized weights \(w^*\) and virtual data (Figure 2). Similarity was then calculated using the weights \(w^*\) at \(q=2\). Japanese women, Japanese men, and foreigners (men) were denoted as 0, 1, and 2, with proportions of 0.1, 0.3, and 0.6, respectively. The network transition from category 0 to 1 and from category 2 to 0 can be visualized, as shown in Figure 2 (the survey network essentially depicts a two-way arrow). Each category proportion is displayed as a circle, and dissimilarities between individuals are displayed as spaces. This visualization aids in intuitively discerning whether a network exists between similar or dissimilar objects based on the relative sizes of the groups. However, in the current context, large positional differences in dissimilarities between groups cannot be effectively represented.

\begin{figure}[!htbp]
    \centering
    \captionsetup{position=below}
    \includegraphics[width=0.8\textwidth]{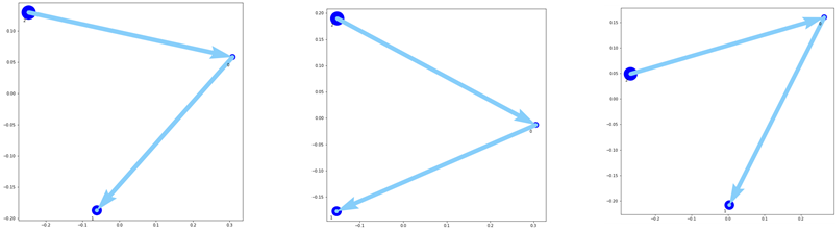} 
    \caption{Figure 2: Examples of visualizations (from left to right: correlation, ordinal correlation, and nonlinear correlation)}
    \label{fig:figure2}
\end{figure}

\section{Discussion}

In this study, a new index that incorporates network and similarity considerations is proposed. First, the properties of the proposed index and its relationship with existing diversity and network indices were investigated. In addition, a method for estimating the weights required for similarity calculations was developed. I also proposed methods for visualization. Finally, the utility of the proposed index was tested using real data. The results show that the proposed DSN method exhibits a stronger correlation than existing diversity indices and network densities, suggesting its potential to accurately represent people's preferences. This suggests the need to examine diversity by considering similarities and networks.

The novelty and distinctiveness of this study are outlined as follows:
\begin{enumerate}
    \item Unlike existing studies, this research enables the consideration of similarity and networks, addressing a form of inclusiveness;
    \item this index encompasses other indices;
    \item it facilitates the estimation of similarity based on utilities; and
    \item the proposed index demonstrates greater utility compared to indices from existing studies.
\end{enumerate}
The efficacy of the proposed index was confirmed through comparisons with indices used in existing studies. These techniques will help in capturing the concepts of ``diversity'' and ``inclusion,'' which are currently focal points in various fields.

Conversely, several challenges persist. In this study, diversity is proposed to be indexed as a scalar, which means that the interactions among proportions, networks, and similarities are accounted for. Each of these processes can be broken down and handled separately. Furthermore, despite partial discussion, the relationships between various types of network diversity indices proposed by Morales et al. and the index in this study have not been thoroughly analyzed \citep{Morales2021}. Further analysis and empirical studies are required, especially in cases where \(E\) is an n-square, as described in the definitions in Section 2.

\section*{Acknowledgement}
This study was supported by JSPS KAKENHI Grant-in-Aid for Scientific Research (C) JP20K02004.

\bibliographystyle{plainnat}
\bibliography{references}

\section*{Appendix}

\textbf{Lemma 1.} \(\sum_{i=1}^{n}\sum_{j=1}^{n}{{(x_i-x_j)}^2}=n\sum_{i=1}^{n}{(x_i-\bar{x})}^2.\)

\begin{proof}The left-hand side is 
\[
\sum_{i=1}^{n}\sum_{j=1}^{n}{\left(x_i-x_j\right)^2}=\left(n-1\right)\sum_{i=1}^{n}{x_i}^2-\sum_{i=1}^{n}\sum_{j=1}^{n}{x_ix_j}.
\]
On the other hand, the right-hand side is 
\[
\sum_{i=1}^{n}{(x_i-\bar{x})}^2=\frac{1}{n}\left((n-1)\sum_{i=1}^{n}{x_i}^2-\sum_{i=1}^{n}\sum_{j=1}^{n}{x_ix_j}\right).
\]
Hence,
\[
\sum_{i=1}^{n}\sum_{j=1}^{n}{{(x_i-x_j)}^2}=n\sum_{i=1}^{n}{(x_i-\bar{x})}^2.
\]
\end{proof}

\end{document}